\documentstyle[pre,aps,multicol,psfig]{revtex}

\begin{document}
\tighten
\draft
\title{Current-induced vortex dynamics in Josephson-junction arrays: Imaging experiments and model simulations.}
\author{S. G. Lachenmann$^\star$, T. Doderer, and R. P. Huebener\\
{\it Lehrstuhl Experimentalphysik II, Universit\"at
T\"ubingen,\\
Auf der Morgenstelle 14, D-72076 T\"ubingen, Germany}}
\vspace{0.9cm}
\author{T. J. Hagenaars$^{\Diamond}$, J. E. van Himbergen,
 and P. H. E. Tiesinga$^{\bullet}$ \\
{\it Instituut voor Theoretische Fysica,\\ Princetonplein 5,
Postbus 80006, 3508 TA Utrecht, The Netherlands}}
\author{Jorge V. Jos\'{e}\\
{\it
Department of Physics,
and Center for
Interdisciplinary
Research on Complex Systems,\\
Northeastern University,\\ 
360 Huntington Ave,
Boston MA 02115, USA}}
\date{\today}

\maketitle
\begin{abstract}
We study the dynamics of current-biased
Josephson-junction arrays with a magnetic penetration
depth $\lambda_\perp$ smaller than the lattice spacing.
We compare the dynamics imaged by low-temperature
scanning electron microscopy to the vortex dynamics obtained from
model calculations based on the resistively-shunted junction
model, in combination with Maxwell's equations.
We find three bias current regions with
fundamentally different array dynamics. The first region is the subcritical
region, i.e. below the array critical current $I_c$.
The second, for currents $I$ above $I_c$, 
is a ``vortex region",
in which the response is 
determined by the vortex degrees of freedom. In this region, the dynamics
is characterized by spatial domains where
vortices and antivortices move across the array in opposite directions
in adjacent rows and by transverse voltage fluctuations.
In the third, for still higher currents,
the dynamics is dominated by coherent-phase motion,
and the current-voltage characteristics are linear.
\end{abstract}
\begin{multicols}{2}
\section{Introduction}

Studies of Josephson-junction arrays 
are of interest to model 
vortex dynamics as well as for their application
in superconducting electronics \cite{TriesteFrascati}.
Most experimental results dealing with (vortex) dynamics in
2D arrays were obtained by 
measuring the current-voltage (I-V) characteristics.
Such measurements do not give the 
spatially resolved information needed for an
unambiguous determination of the detailed
microscopic dynamics underlying the measured
response.
By contrast the 
low-temperature scanning electron microscopy (LTSEM)
is a technique that 
allows for spatially resolved investigation
of the dynamical states in superconducting systems.
As such it offers the possibility to 
determine the microscopic nature
of the dynamics, if one correlates the experimental information
with the microscopic dynamics obtained 
from model calculations.

Here we present both experimental images and theoretical 
vortex dynamics results
in dc-biased 2D classical arrays (i.e. with a Josephson coupling
energy $E_J$ much larger than the charging energy $E_c$) with high
damping and a small magnetic penetration depth, and in zero
applied magnetic field.
We gain detailed insight
in the spatially resolved vortex dynamics
by supplementing and comparing the  LTSEM results to
those obtained from model calculations employing the 
resistively-shunted junction equations together
with Maxwell's equations.
Due to the small magnetic penetration depth of our samples,  
the applied dc current gives rise to strong induced magnetic fields
at the edges of the arrays. These fields in turn  
facilitate the penetration of (current-induced) vortices at the edges.
In this respect there is some correspondence between
inductive overdamped arrays and
continuous superconducting thin films.
Both systems show a
current-induced resistive state due to the 
nucleation of vortices of opposite
vorticity at opposite
ends of the sample and subsequent vortex
motion into the sample, as is described for superconducting
bridges in References
\cite{Asl,Hue76}.

We include the mutual inductances
between array cells
in order to take
into account the self-induced fields in the model simulation.
We use an algorithm developed recently in
Ref. \cite{Dom94}, that
takes into account an approximate full-range inductance matrix.
A particularly interesting region of array dynamics is the current
region
slightly above the array critical current. There we find an
intricate structure in the I-V curves 
\cite{Lac94,Lac95a,Dod95a,Dod95b}.
Recent results of LTSEM experiments  in this current region
were interpreted in terms of
the collective motion of current-induced vortices \cite{Lac94}. 
Subsequently the detailed form of this motion was  found to be in close
agreement with preliminary results
of numerical investigations, that take into account
inductive effects \cite{Hag95}.

The aim of this work is to provide insight
into the array dynamics underlying the 
structures found in the I-V curves over the whole
current range. To achieve this goal we compare the experimental
imaging results to graphical animations of the
time-evolution of the spatially resolved vortex pattern
distributions obtained from our model calculations.
In addition, we define and calculate a number of
order-parameter-like quantities that characterize the 
nature of the microscopic dynamics.
The main conclusion is that we can distinguish three different
regions in the array dynamics: the subcritical region (I), the vortex
region (II), and the region of constant differential resistance (III). In 
region (I) the array is in a zero-voltage state.
Region (II) is dominated by (collective) vortex
dynamics, 
contributing to the structure of the I-V characteristics.
In contrast, in (III) the
array dynamics is characterized by a
`nearly coherent' behavior of the junction phases.
In Section II we describe the samples and the imaging technique.
In Section III 
we introduce the model equations and the quantities calculated.
In Section IV the
experimental measurements and the model calculation
are discussed and compared.
Our conclusions are presented in Section V.

\section{Experimental Techniques}
\subsection{Samples}

The samples used for the present studies consist of two-dimensional
arrays of ${\rm Nb/AlO_x/Nb}$ junctions with
square elementary cells. The junctions are square with an
area of about ${\rm 18 \ \mu m ^2}$.
We used 6$\times$6, 10$\times$10,
and 20$\times$20 arrays without a ground plane and
10$\times$10 and 20$\times$10 arrays with a superconducting
PbIn ground plane placed at a distance to the array
of about 1 ${\rm \mu m}$. Here an $N \times M$ array
denotes an array that has $N$ columns of $M$ junctions.  
The lattice spacing $a$ is 16.7 ${\rm \mu m}$.
Each of the junctions is externally shunted by an ohmic resistor
${R_s \approx 1.5 \ {\rm \Omega}}$ to decrease the McCumber parameter
$\beta_c =2\pi i_c R_s^2 C / \Phi_0 \approx 0.7$\cite{Ben91}
(overdamped regime),
where
${ \Phi_0 = h/(2e)}$ denotes the flux quantum.
The critical current of each junction is $i_c \approx 150 \ {\rm \mu A}$.
The spread of $i_c$ over one array is typically less than
3\% (one standard deviation from the mean value)\cite{Ben91}.
The magnetic penetration depth ${ \lambda_{\perp}
= \hbar /(2e\mu _0 i_c)}$ of the arrays \cite{Lik}
is smaller than $a$, where ${ \mu _0}$ is the
permeability of free space and $e$ is the elementary charge. 
In Ref.\cite{Ben91} the
sample geometry, layout, and fabrication is described in more
detail.

\subsection{Experimental Imaging of Arrays}

Low-temperature scanning electron microscopy 
offers the possibility to image various
properties of superconducting
samples during their operation at liquid helium temperatures.
The basic LTSEM principles together with some
results are described in Refs.~\cite{Hue88,Gro}.
The top surface of the sample is scanned
with the electron beam, while the sample is thermally
coupled to a liquid helium bath.
For the present studies, the sample is dc current biased
and the electron beam induces    a change
${ \Delta V}$ in the array voltage 
that is recorded as a function of the focus
coordinates ${ (x_0,y_0)}$ of the e-beam.
In order to increase the sensitivity, the e-beam is chopped with a
20 kHz frequency and ${ \Delta V}$
is phase-sensitively detected with a lock-in amplifier.
Typical values for ${ \Delta V}$ are in the the range
100 nV -- ${\rm 5 \ \mu V}$, whereas the array voltage, $V$,
is of the order of
mV. Hence, the perturbation due to the e-beam irradiation
is small.
The sample
temperature is estimated to be about 4.5 K for a
Helium-bath temperature of 4.2 K.
The dominant effect of the e-beam irradiation
is local heating. We estimate from the e-beam parameters
(electron energy: 25 keV, beam current:
100 pA) that there is a local temperature increment at the
beam focus
of about 0.4 K. The lateral extension
of the
thermally perturbed area is about 1 ${\rm \mu m}<a$
(representing the limit of the spatial resolution
of this imaging technique).

The sample is shielded from external magnetic fields
by four ${\rm \mu}$-metal shields at both room and
liquid helium temperatures. 
A perpendicular magnetic field can be
applied using a circular copper coil placed
in the liquid helium just below the sample substrate.
In zero applied magnetic field,
the residual external dc field perpendicular to the array,
corresponds to a frustration
${ f \lesssim 0.1}$, 
with $f$ the average external
flux in one unit cell divided by ${\Phi _0}$.
This residual magnetic field
${ B \approx 700}$ nT, was obtained
by measuring the dependence of the array critical current
${ I_c}$ on an applied perpendicular magnetic field
for an array without a superconducting ground plane.

The experimental data shown in this paper is obtained
in zero applied perpendicular magnetic field, i.e.
in the residual magnetic field mentioned above.
For applied fields corresponding
to approximately integer values of $f$ the 
experimental results remain qualitatively the same up
to ${f \approx 5}$.

To interpret our imaging results, it is important
to identify the different time scales involved.
The junction oscillation period is of the order of 10 ps,
whereas the decay time of the beam's thermal perturbation is about
100 ns\cite{Hue88}.
During step-wise scanning, the electron beam typically stays 3 ms at
each position.
The time needed to take one complete
LTSEM image of the array dynamics is of the order of
minutes, hence, the beam-induced voltage signal
$\Delta V (x_0,y_0)$ represents a {\em time-averaged}
quantity on the time scales of Josephson dynamics.

The local temperature increment at the e-beam focus
is most effective at the positions of the Josephson junctions.
For an e-beam current of
100 pA, the heating of an individual junction at $(x_0,y_0)$
results in a reduction
$\Delta i_c$ of the critical current of this junction
$\Delta i_c / i_c \approx 8 \%$.
The resulting voltage change $\Delta V(x_0,y_0)$ depends on the nature
of the dynamics at and around the junction at $(x_0,y_0)$.
Hence, a spatially inhomogeneous steady-state dynamics gives
rise to an inhomogeneous image.
Spatially resolved images have been interpreted in Ref. 
\cite{Lac94} in terms of vortex motion or, in the case
of underdamped samples, in terms of row-switched
dynamical states.

\section{Model equations for the dynamics of inductive Josephson-junction arrays}
In this section we briefly discuss the 
model we use to describe the array dynamics including the 
self-induced magnetic fields.
In the model, the array is driven by a
uniform applied dc current ${I}$
along the vertical direction.
In the classical regime,
$E_{J} \gg E_{c}$, ($E_J = \Phi_0 i_c / 2 \pi$ and $E_c =e^2/2C$),
the phases $\theta({\bf r})$ 
of the superconducting order parameter on an island ${\bf r}$,
are the only
variables. 
The array dynamics is then determined by the resistively
and capacitively shunted junction (RCSJ) model for each 
junction, Kirchhoff's
current conservation condition on the superconducting islands, 
plus Faraday's law for the magnetic field dynamics.

Using the approximation introduced
in Ref. \cite{Dom94} (``Model C")
for the full-range inductance matrix,
and using the temporal gauge, one obtains a closed set of
dynamical equations for the gauge-invariant phase differences
$\Psi({\bf r},{\bf r'})\equiv
\theta({\bf r})-\theta({\bf r'})-2\pi A(\bf{r},\bf{r'})$.
Here $A(\bf{r},\bf{r'})$ is defined
by the line integral of the vector potential $\bf{A}$,
$A({\bf r},{\bf r'})=(1/\Phi_0)\int_{\bf r}^{\bf r'}{\bf A}\cdot d{\bf l}$.
The derivation and implementation of these model equations 
is discussed in more detail in Refs. \cite{Dom94,Dom96}. In the
model calculations we can
explicitly tune the dimensionless parameters $\beta_c$ and
$\lambda_\perp$, introduced in Section II.A.\\
In our simulations we obtain $\Psi({\bf r},{\bf r+a})$ (${\bf
a}={\bf e}_x,~{\bf e}_y$) as a function of time.
The magnetic flux is then given by
\begin{equation}
\frac{2\pi\Phi({\bf R },t)}{\Phi_0}=\sum_{{\cal P}({\bf R })} \Psi({\bf r },{\bf r' },t).
\label{goaround2}
\end{equation}
Here ${\cal P}({\bf R})$ is the anti-clockwise
sum over the four bonds $({\bf r},{\bf r'})$ around the plaquette
with coordinate ${\bf R}$. 
The vorticity $n({\bf R})$ is given by 
\begin{equation}
2\pi (n({\bf R})-\Phi({\bf R})/\Phi_0)=-\sum_{{\cal P}({\bf R})} {
\tilde{\Psi}({\bf r},{\bf r'})},
\end{equation}
with
\[
\tilde{\Psi}({\bf r},{\bf r'})
=\Psi({\bf r},{\bf r'})- 2\pi\, {\cal N} \left(\frac{{\Psi}({\bf r},{\bf
r'})}{2\pi} \right),
\]
or, equivalently, by
\begin{equation}
n({\bf R},t)=-\sum_{{\cal P}({\bf R})} {\cal N} 
\left(\frac{\Psi({\bf r},{\bf r'})}{2\pi} \right).
\end{equation}
Here the function $\cal N$ yields the integer nearest to 
the argument.
The voltage response for a current applied in the $y$-direction
is obtained from
\begin{equation}
V=\frac{1}{N M N_t} \sum_{\bf r} \sum_{t=1}^{N_t} \frac{d\Psi({\bf r},{\bf r-e}_y,t)}{dt}.
\label{voltind}
\end{equation}
Here 
$N_t$ is the number of
time-integration steps, $\sum_{\bf r}$ is the sum over the
$N\times M$ junctions in the current direction.
 $V$ is expressed in units of $i_c R_s$,
time is expressed in units of the dimensionless characteristic time 
$t_c=1/\omega_{c}=\hbar/(2eR_{s}i_{c})$.

\section{Experimental and theoretical results}
\subsection{Current-Voltage Characteristics}
Figure~1
shows a representative experimental current-voltage (I-V) characteristic 
together with its differential resistance $dV/dI$
obtained from a 10$\times$10 array.
The rich structure of $dV/dI$
above $I_c$
is typical for all arrays studied.
In 
Fig.~1
we indicate three regions. The
subcritical region (I) defined for $I < I_c$, an intermediate
current region (II)
ending at $I_{lin}$, and region (III), where the 
differential resistance becomes constant, for
$I > I_{lin}$.
\begin{figure}
\centerline{
\psfig{figure=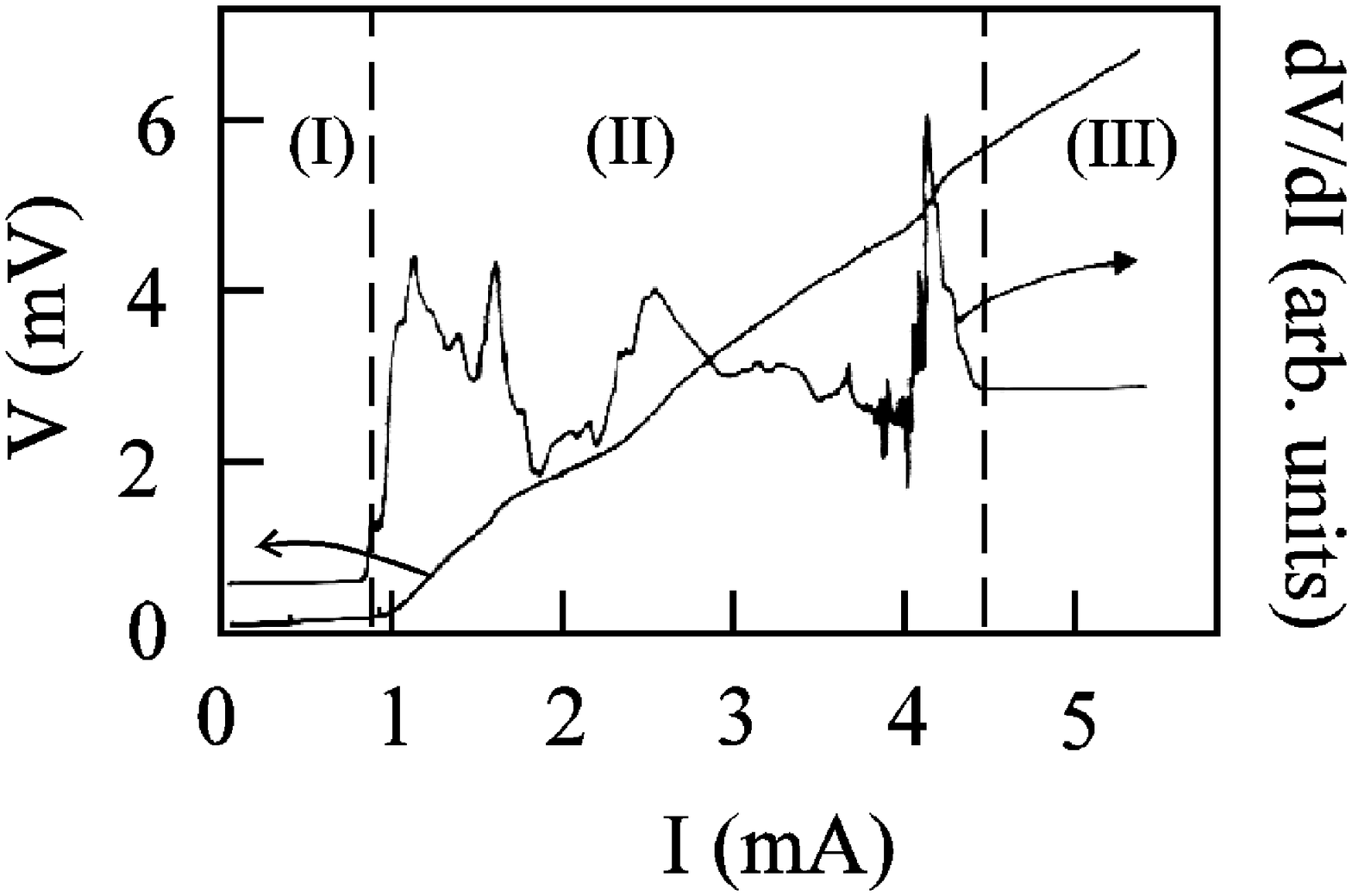,height=7.0cm,width=8.0cm}
}
{\small FIG.~1.
I-V characteristic of a
10$\times$10 array together with the differential
resistance $(dV/dI)(I)$
measured at a temperature $T\approx 4.5$ K.
\label{IVCex} 
}
\end{figure}
In Fig.~2 
we show an I-V curve obtained from numerical
simulation of a
$20\times 20$ array with $\beta_c=2\pi i_c R_s^2 C / \Phi_0
=0$ and $\lambda_\perp =0.6a$.
Also shown is $dV/dI$ (thin line).
The critical current
is  $I_c/N=0.87i_c$.
This I-V curve is qualitatively similar to the one measured in the
experiments:
for intermediate currents, the differential resistance
shows a jagged structure, and
for currents $I \gtrsim 1.5Ni_c$ 
the I-V curve is linear, just
as is the case for large currents in the experimental I-V curve.
As our prime goal is to obtain
a qualitative modeling of the
experimental systems,
the choice of simulation parameters is motivated partly
by numerical convenience.
In particular, we simulate a $\lambda_\perp=0.6a$, although
in experiment this ratio is about $0.1$. 
The latter value would lead to considerably
stiffer differential equations that require much longer 
computation times.
Due to their smaller $\lambda_\perp/a$ ratio, 
the samples studied experimentally have 
even stronger self-induced magnetic
fields than the simulated ones, and thus
smaller critical currents.
Furthermore, in experiment, region (II)
extends up to $I_{lin}\approx 2.9 Ni_c$, whereas
the simulations indicate that the border between regions
(II) and (III) is at $I\approx1.5Ni_c$. 

\begin{figure}
\centerline{
\psfig{figure=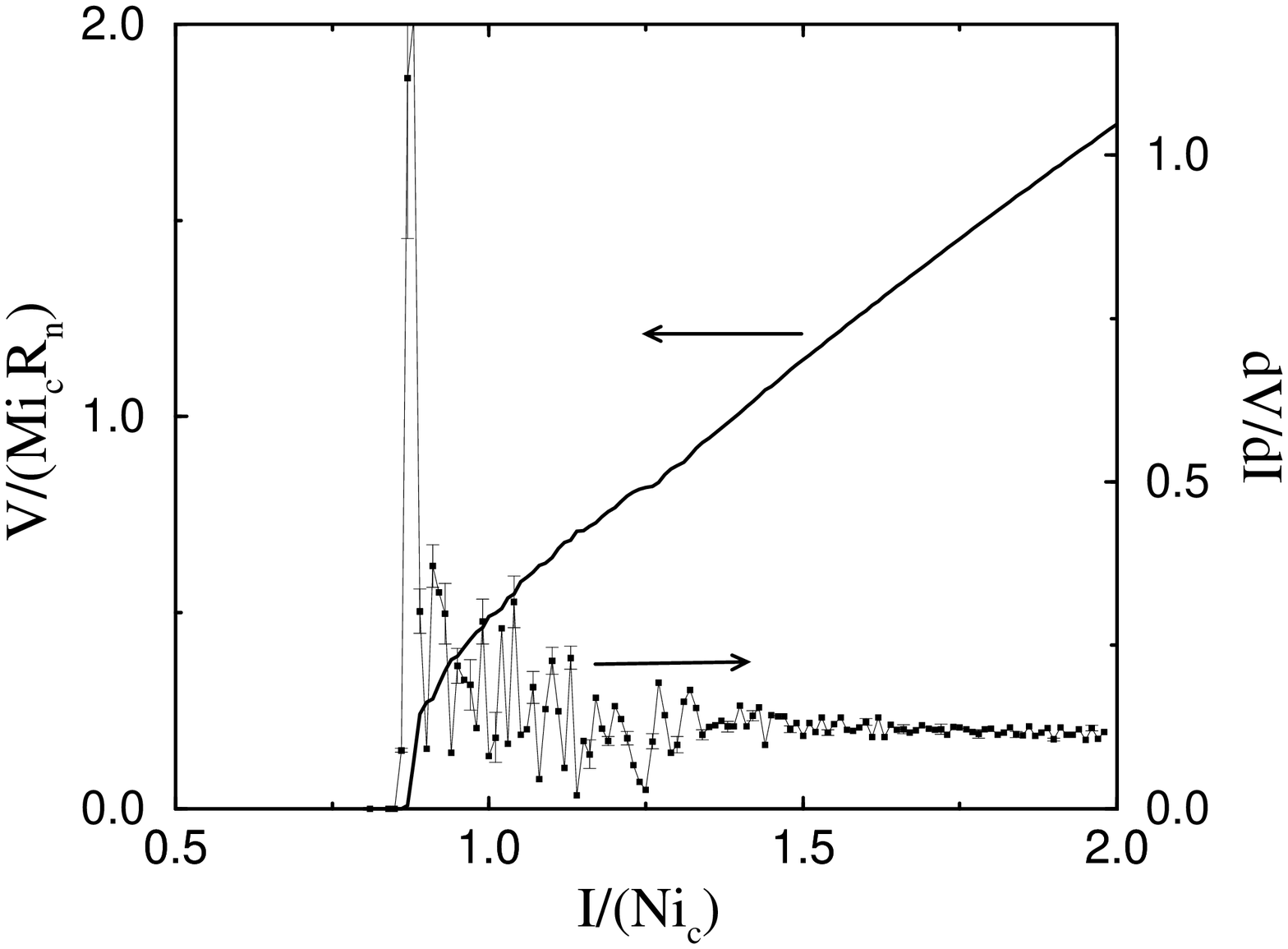,height=7.0cm,width=8.0cm}
}
{\small FIG.~2.
Theoretical zero temperature $I-V$-characteristics
(fat line)
and the differential resistance
$(dV/dI)(I)$ (for selected data points we also show
the error bars)
of a 20$\times$20 array
with $\beta_c=0$, and $\lambda_\perp =0.6a$.
$V$ is the time-averaged voltage across the array in the current
direction.} 
\label{IVCth} 
\end{figure}

\subsection{Imaging of Array Dynamics}

In this section we present
images of the array dynamics for the three 
different regions in 
the I-V characteristic 
in Fig.~1. We first discuss the subcritical region, then the
region of vortex dynamics, and finally the region of 
constant differential resistance.
Typical LTSEM imaging results are shown in Fig.$~3$
and model simulation results  in Fig.$~4$.
\subsubsection{Subcritical region (${I<I_c}$)}
For a bias current $I < I_c$ and in the absence of LTSEM heating,
the array is in the zero-voltage state.
Nevertheless we obtain useful information
from the dynamical imaging experiments.
In particular, we will see below that
the LTSEM images obtained in this region
confirm the importance 
of inductive effects in our small-$\lambda_\perp$ samples.

In Figure 3 (a) we show a typical LTSEM imaging result
below but close to the array critical current
$I_c$, for relatively high 
beam power.
From this result we see
that junctions at or near the
edges parallel to the bias current
give a voltage response
to the local heating.
This can be understood in the following way.
For currents in this subcritical region,  but close to the array
critical current $I_c$, the LTSEM is acting as an active probe,
inducing vortex motion.
When  the $i_c$ of a junction at the array
edge is lowered due to the e-beam irradiation, vortices can overcome
the energy barrier for entry 
at this junction and subsequently travel across
the array. As a result, a voltage signal
$\Delta V > 0$ is observed. 
The corresponding process for junctions not at the
edges is the creation of a vortex-antivortex pair.
From Fig. 3(a) we see that this latter process
does not occur for junctions that are not close to the edges.
This means that such junctions carry less current, which
corresponds to larger energy barriers for vortex-antivortex creation.
Such an inhomogeneous distribution of the bias current
is in agreement to the small magnetic penetration depth
$\lambda$ of the array.
The correspondingly strong inductive effects
lead to a spatial distribution of the dc bias current 
that is strongly peaked at the two array edges
parallel to the current flow
\cite{Phi94,Bus94}.
The current-induced  flux is also maximal at these edges.
It is oriented in opposite direction at
opposite edges. 
This is illustrated in Fig.~5, where we plot the numerically calculated
current and magnetic field distributions for a $20 \times 20$ array
at $I=0.86Ni_c$ and with $\lambda_\perp=0.6a$.
Similar simulation results in this subcritical region have
been reported in Refs. \cite{Dom96,Phi94}.

\subsubsection{Vortex dynamics region ($I_c < I < I_{lin}$)}
Above $I_c$
the current through the edge junctions  exceeds the junction critical
current. As a result, 
vortices enter the array at one edge and antivortices at the opposite edge.
These vortices are depinned from the edges
by the Lorentz force and
move across the array, generating the observed
voltage.

LTSEM results for bias currents slightly above
$I_c$ are presented and discussed in Refs.\cite{Lac94,Lac95a,Dod95a,Dod95b}.
In the LTSEM images obtained in this current region
it is seen (see, e.g., Ref.~\cite{Lac94}) that
the sign of the voltage signal
near the sample edges tends to alternate
along the current direction.
These images indicate an alternating or staggered crossing vortex motion
where vortices and antivortices are 
nucleated at opposite
array edges and subsequently move across the whole array.
At the array edge opposite to the nucleation site
they leave the array or, equivalently,
annihilate with an image vortex of opposite sign.
\end{multicols}
\newpage
\begin{figure}
\centerline{\hspace{1.0cm} (a) \hspace{8.5cm} (b) }
\centerline{}
\centerline{
\psfig{figure=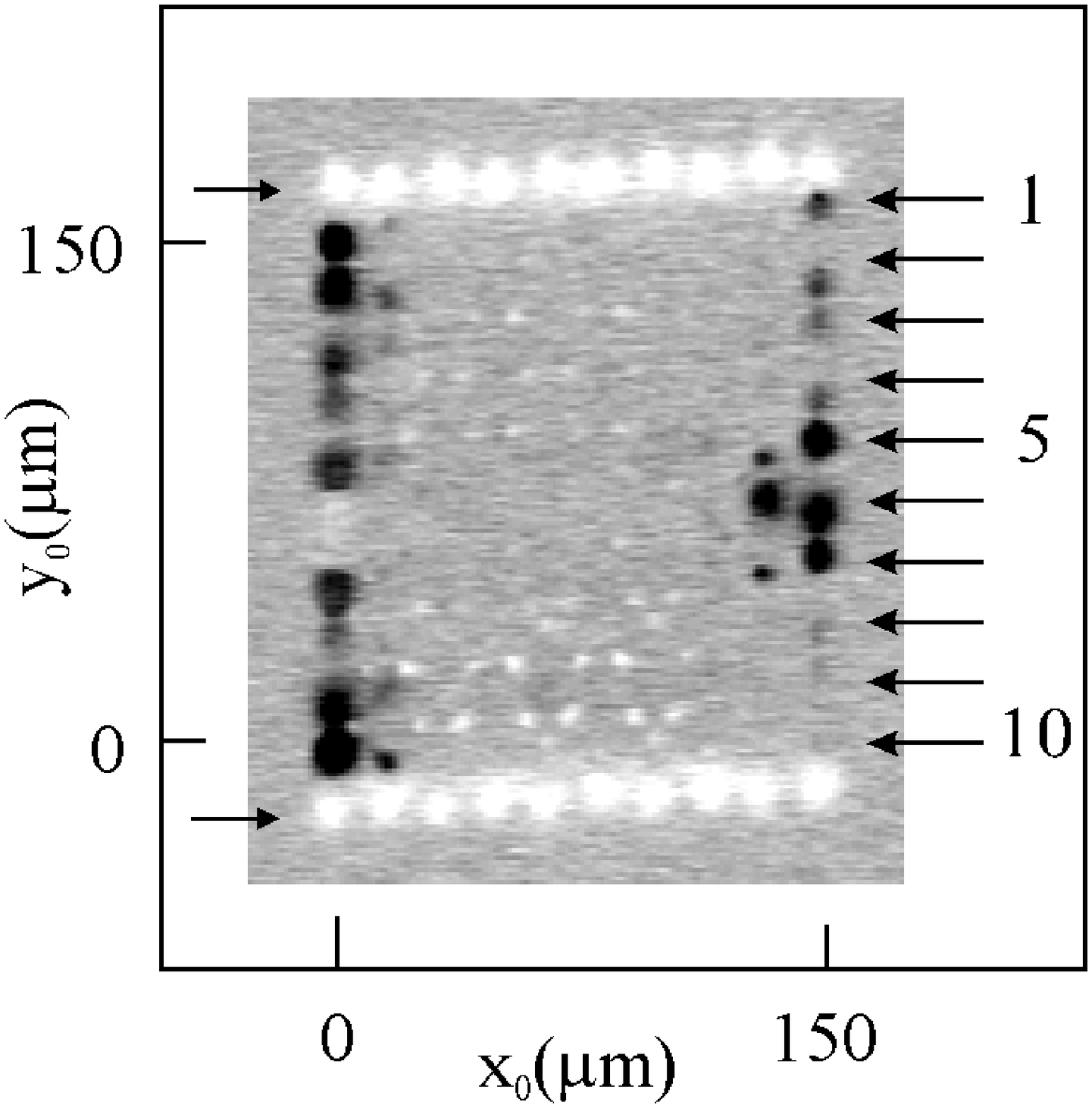,height=6.7cm,width=8.0cm}
\hspace{1.6cm}
\psfig{figure=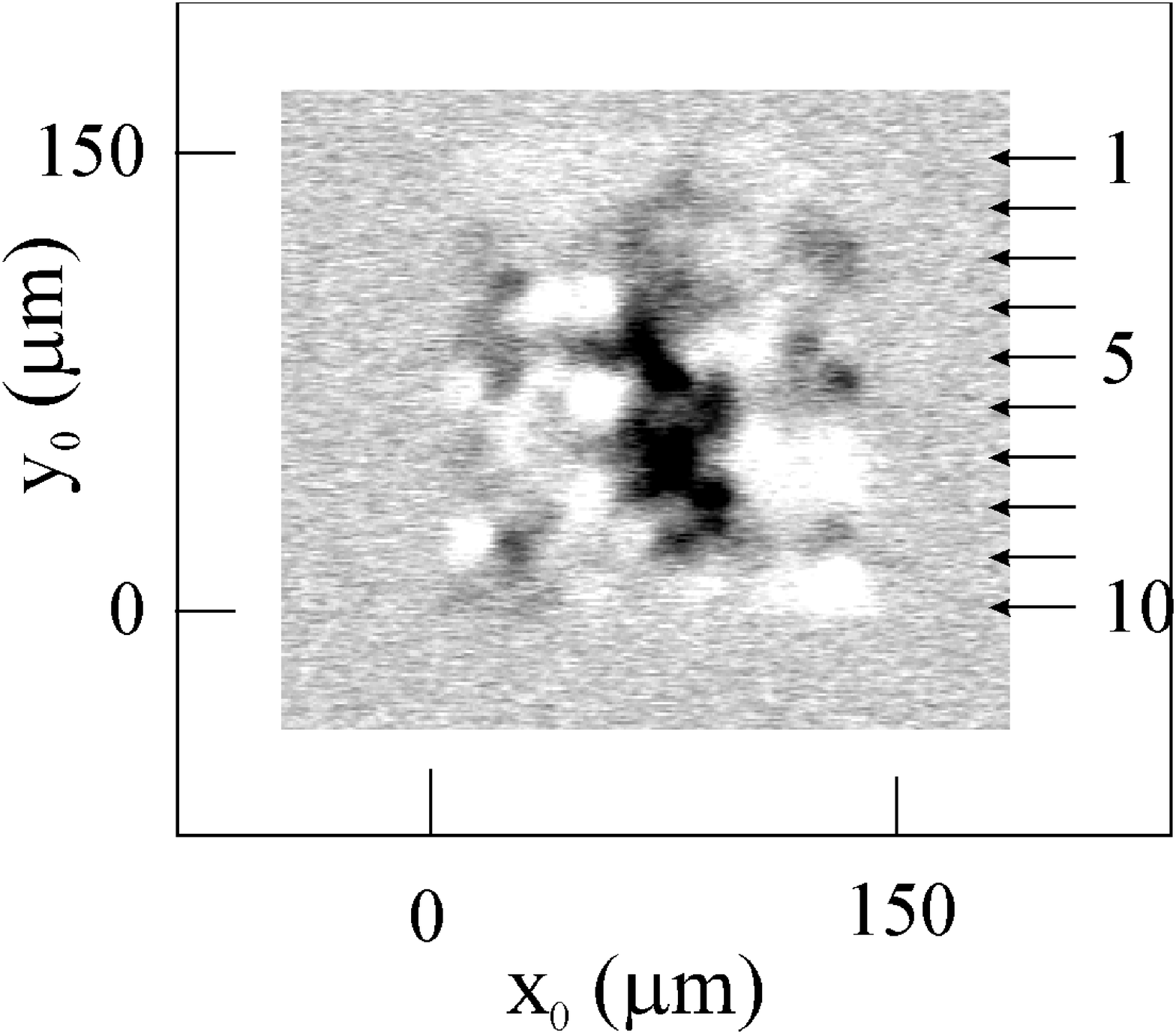,height=6.7cm,width=7.0cm}
}
{\small FIG.~3.
Grey value representations of the experimental 
voltage image $\Delta V(x_0,y_0)$ for a 10$\times$10 array at $T\approx 4.5$ K.
The array is current biased at $I=0.9 I_c$ (a) and $I=4.25 I_c$ (b),
respectively.
The dc bias current flows vertically through the array. The array
boundaries lie between 0 ${\rm \mu m}$ and 150 ${\rm \mu m}$
in both directions. A positive (negative) e-beam induced
voltage signal $\Delta V(x_0,y_0)$ is indicated by the dark
(bright) areas, whereas zero signal is shown
by the area surrounding the array.
The individual rows of junctions are indicated
by the small arrows numbered 1--10 from top to bottom.
In (a), the $\Delta V < 0$ voltage response 
at the top and bottom of the array columns (marked by arrows
from the left)
arises from the current feeding resistors 
made from InAu thin films.}
\label{LTSEMsub}
\label{LTSEMins} 
\end{figure}
\begin{figure}
\centerline{(a) \hspace{5.5cm} (b) \hspace{5.5cm} (c)}
\centerline{
\psfig{figure=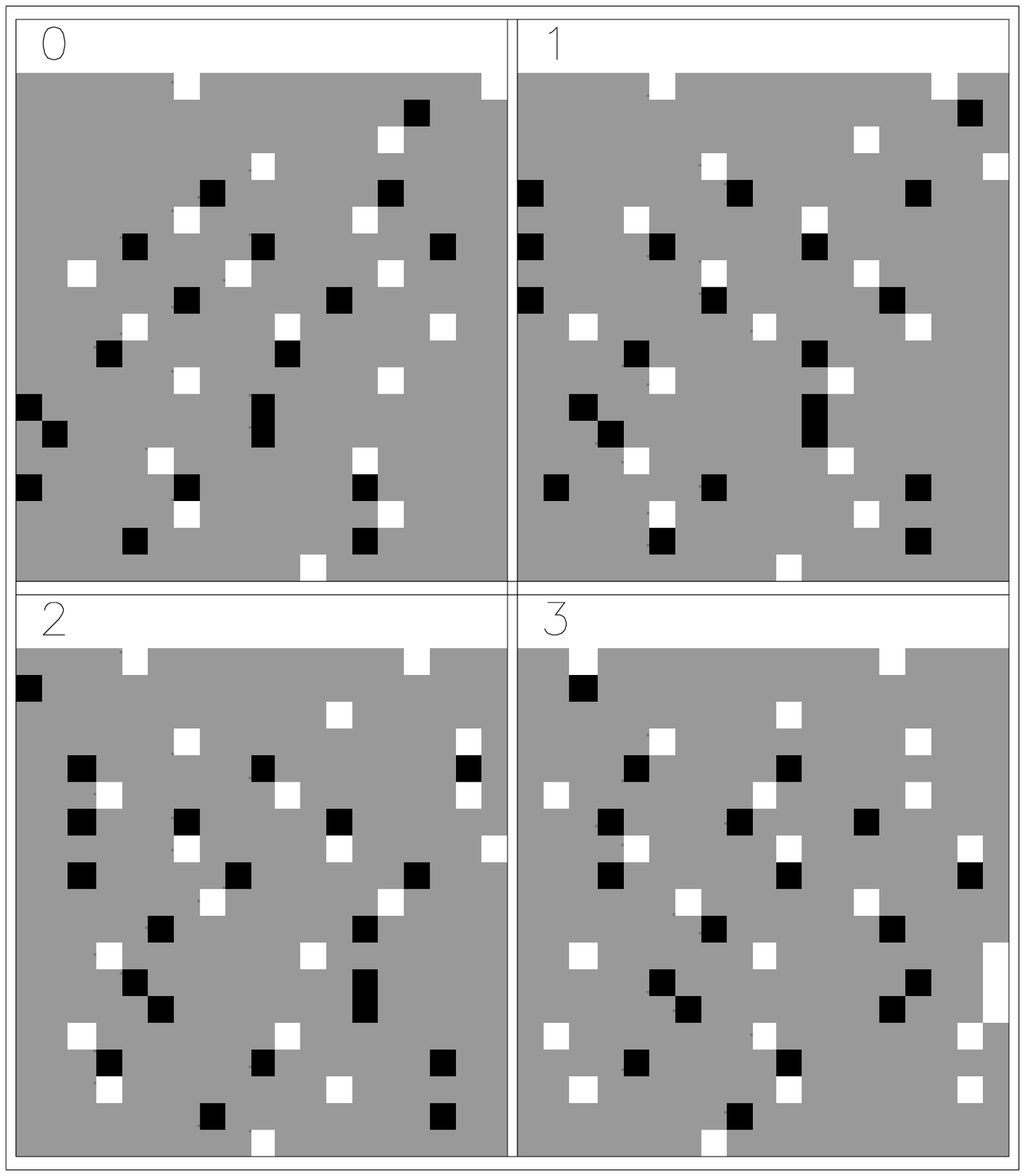,height=5.5cm,width=5.5cm}
\hspace{0.5cm}
\psfig{figure=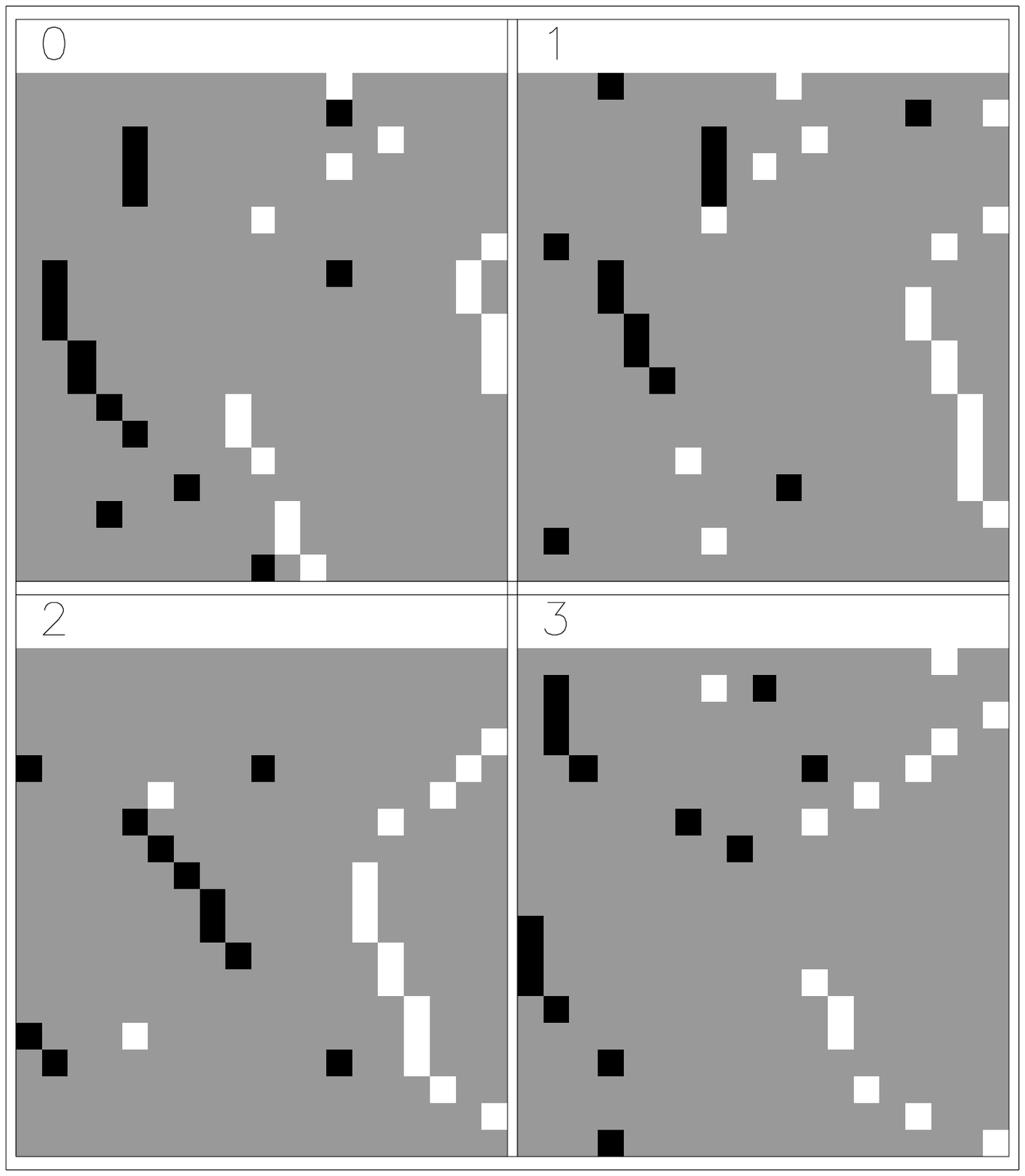,height=5.5cm,width=5.5cm}
\hspace{0.5cm}
\psfig{figure=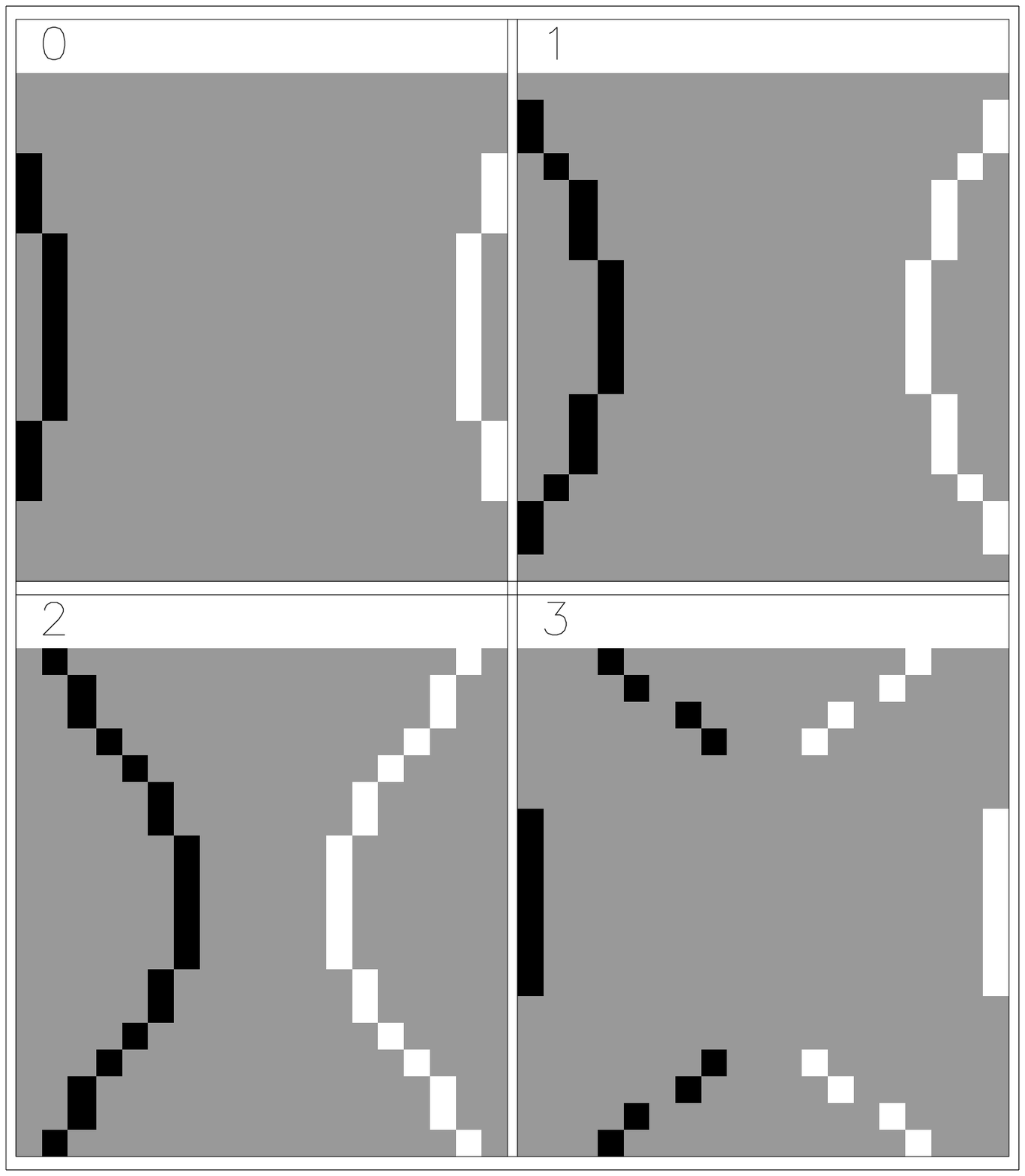,height=5.5cm,width=5.5cm}
}
\centerline{}
{\small FIG.~4. 
Vortex configurations
for simulations of a 20$\times$20 array, for
three different values of the bias current (a-c).
For each current value, we show four consecutive
frames (0-3).
The dc bias current flows vertically through the array.
The black (white) squares
denote plaquettes with vortex number $n({\bf R})=+1(-1)$.
Vortices with positive sign move to the right, vortices
with negative sign to the left, as can be deduced by comparing
consecutive frames.
(a) $I=0.95Ni_c$.
Frame $0$ is at $t=3000t_c$. Between consecutive frames
there is a time interval of $3t_c$.
Adjacent rows tend to be crossed by vortices of opposite
sign. The alternating structure is disrupted between
the third and fourth row from above and between the
sixth and seventh row from below.
(b) $I=1.20Ni_c$.
Frame $0$ is at $t=9025t_c$. Between consecutive frames
there is a time interval of $2t_c$.
(c) $I=2.0Ni_c$. Frame $0$ is at $t=500.75t_c$. Between consecutive frames
there is a time interval of $0.9t_c$.
}
\label{vortexP200}
\end{figure}
\newpage
\begin{multicols}{2}
\noindent
We will see below that this interpretation is  confirmed  by 
the general behavior found in simulation images
for a bias current slightly above the array critical current.
This kind of alternating crossing vortex motion
has no  observed analog in continuous superconducting samples.

To analyze the dynamics of the vortices in full
detail, we have 
studied graphical animations of the time evolution of the vortex
distributions in the array.
First we discuss the ones obtained in a
simulation for $I=0.95Ni_c$
of a $20\times 20$ array with $\beta_c=0$ and $\lambda_\perp
=0.6a$. When we start the simulation with random initial
phases
we observe, 
after a transient of about $t/t_c=800$ time units,
the type of vortex
motion depicted in 
Fig.~4 (a).
The snapshots in 
Fig.~4 (a)
show the same type of vortex
motion as the one deduced from the
LTSEM measurements:
vortices of opposite sign cross the array in opposite directions in
adjacent rows.
We observe that the staggered  structure is broken at two
places, where two adjacent rows are crossed by vortices of equal
sign. One might view these places as domain
wall defects between two different polarities of the staggered pattern
(for experimental results
see also 
Ref.~\cite{Lac94}).
We have verified that this type of dynamics is stable for
very long simulation times. 
The position and number of domain walls
depends sensitively on the initial conditions.
For all currents in the range
$I_c/(Ni_c)=0.87\leq I/(Ni_c)\lesssim 1.15$
the long-time stable vortex dynamics is of the same staggered type. 
\begin{figure}
\centerline{
\hspace{-1.0cm}
\psfig{figure=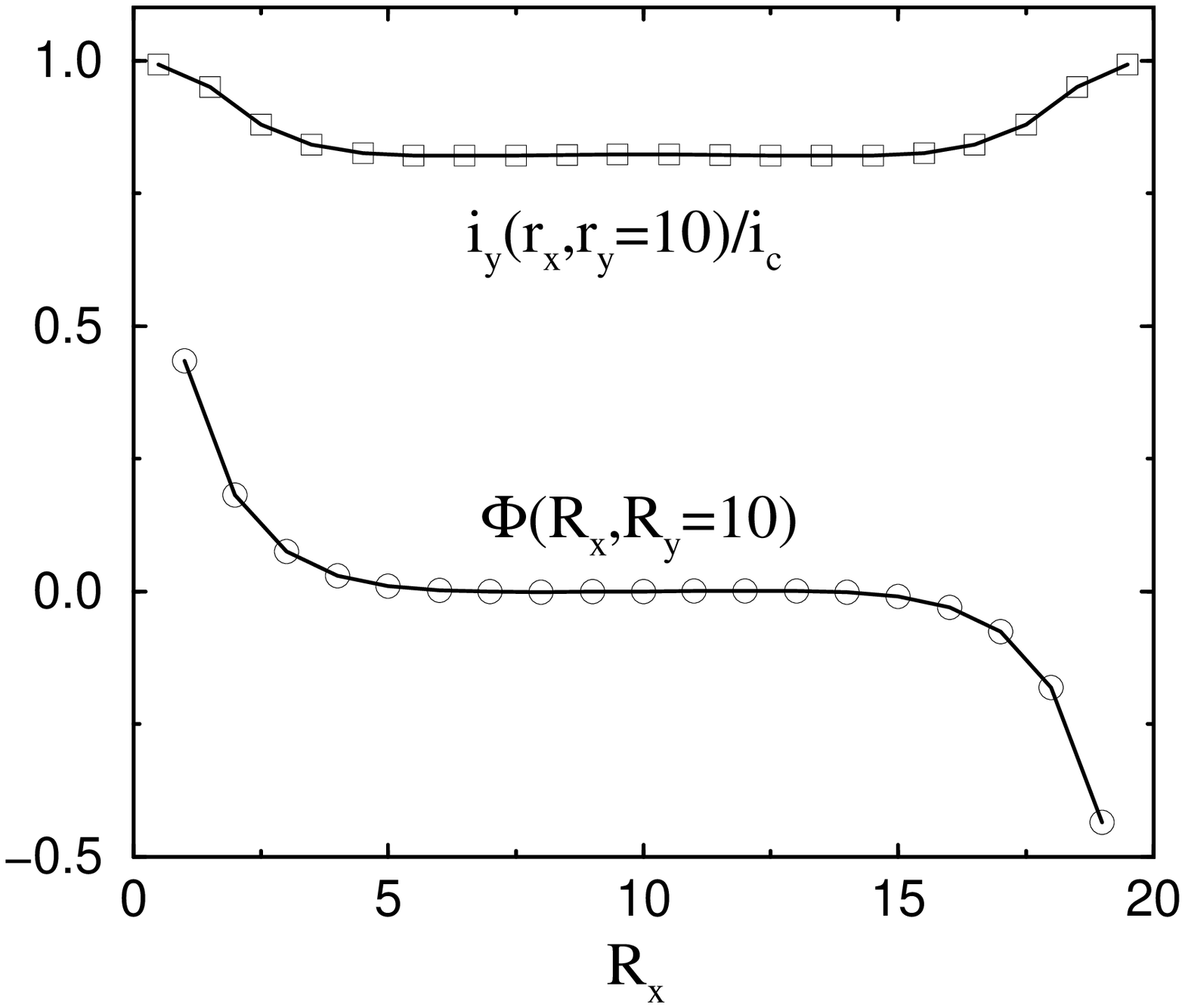,height=6cm,width=7.0cm}
}
{\small FIG.~5. 
Simulated 
field and current distribution in the central row of a $20\times 20$ array with. $\lambda_\perp=0.6a$ and $I=0.86Ni_c$, just below the array
critical current. $i_y$ is the current through a longitudinal junction.} 
\label{edgefields} 
\end{figure}
The dynamical patterns observed experimentally as
well as numerically for $I\gtrsim I_c$ 
are strongly influenced by vortex-vortex
interaction.
These interactions lead to
the almost regular  patterns in which 
the vortices tend to move.
Towards the high current end of region (II), the
dynamics is different.
Imaging the dynamical state by LTSEM 
now yields larger two-dimensional domain
patterns spreading over
several unit cells in both the $x$ and $y$-directions. 
A typical example for this kind of voltage response is given
in Fig.~3 (b).
    From our experimental
observations, we can deduce the following: (1) The dynamical state, which gives
rise to the voltage image is stable in time.
If the parameters of the sample ($I$, $T$) are not changed a subsequent
image will give the same result.
(2) The imaging results depend sensitivily
on the history. If, for example,
the bias current or temperature is changed significantly, and then
returned to the same parameter values, the imaging results
change.
(3) When $I$ or $T$ is changed smoothly, we observe a smooth variation of
the detected patterns.
(4) A magnetic field (noninteger
$f$ or $|f| >5$) changes the regular pattern,
observed for small bias currents discussed above,
to a complex response similar to that shown in 
Fig.~3 (b).
Based on these observations, we conclude that the
voltage response is caused by a complex multi-vortex dynamics, and
not by e.g.\ sample inhomogeneities, trapped flux, or
temperature fluctuations.

In our simulations, the staggered vortex dynamics
is the relevant dynamics up to approximately
$I=1.15Ni_c$. The region $1.15 \lesssim I/(Ni_c) \lesssim 1.5$ is a transition region
between the regime of staggered vortex dynamics and the regime of
constant differential resistance.
In this current interval,
in some parts of the array vortices
move independently,
and in  others we observe vortices that
tend to move
coherently in adjacent rows.
A representative example  of this type of dynamics is given in
Fig.~4 (b).

\subsubsection{Linear branch ($I > I_{lin}$)}
In region (III), where the I-V curve is linear (constant
differential resistance), each junction
in the array columns (longitudinal junction)
yields approximately the same
voltage signal $\Delta V (x_0,y_0)$
and the LTSEM image
is rather uniform over the whole array. For example, at
$I = 5 I_c$ for a $10 \times 10$ array the voltage signal at
each junction was the same within $5\%$ (see Fig.~7 of \protect{\cite{Dod95b}}).

The absence of any structure in the
beam-induced voltage signal
suggests that there are no isolated vortices entering or
leaving the array.
This is  indeed what we find in  the simulations in
the region of constant differential resistance.
In fact, the dynamics  in this current region 
is not due to vortex motion but to a
state in which the longitudinal junctions belonging to the same
column oscillate almost in phase.
This phase coherence reveals itself in a wave-like dynamics
of the magnetic field distribution.
When looking at the discrete vortex configurations shown in
Fig.~4 (c),
we observe fronts  of vortices that move inward from the
boundaries. In the middle the 
vortices annihilate.
The discrete vortex configurations
in Fig.~4 (c) show
a high degree of symmetry,
which reflects the coherence in the motion of the
longitudinal junctions
in different rows.
We have also explicitly simulated a scanning-induced  8\% critical-current
reduction of subsequent individual junctions for $I=2.0Ni_c$, and indeed find a
spatially uniform voltage change. 
\newpage
\end{multicols}
\begin{figure}
\unitlength=0.1in
\begin{picture}(60,30)(0,0)
\includegraphics{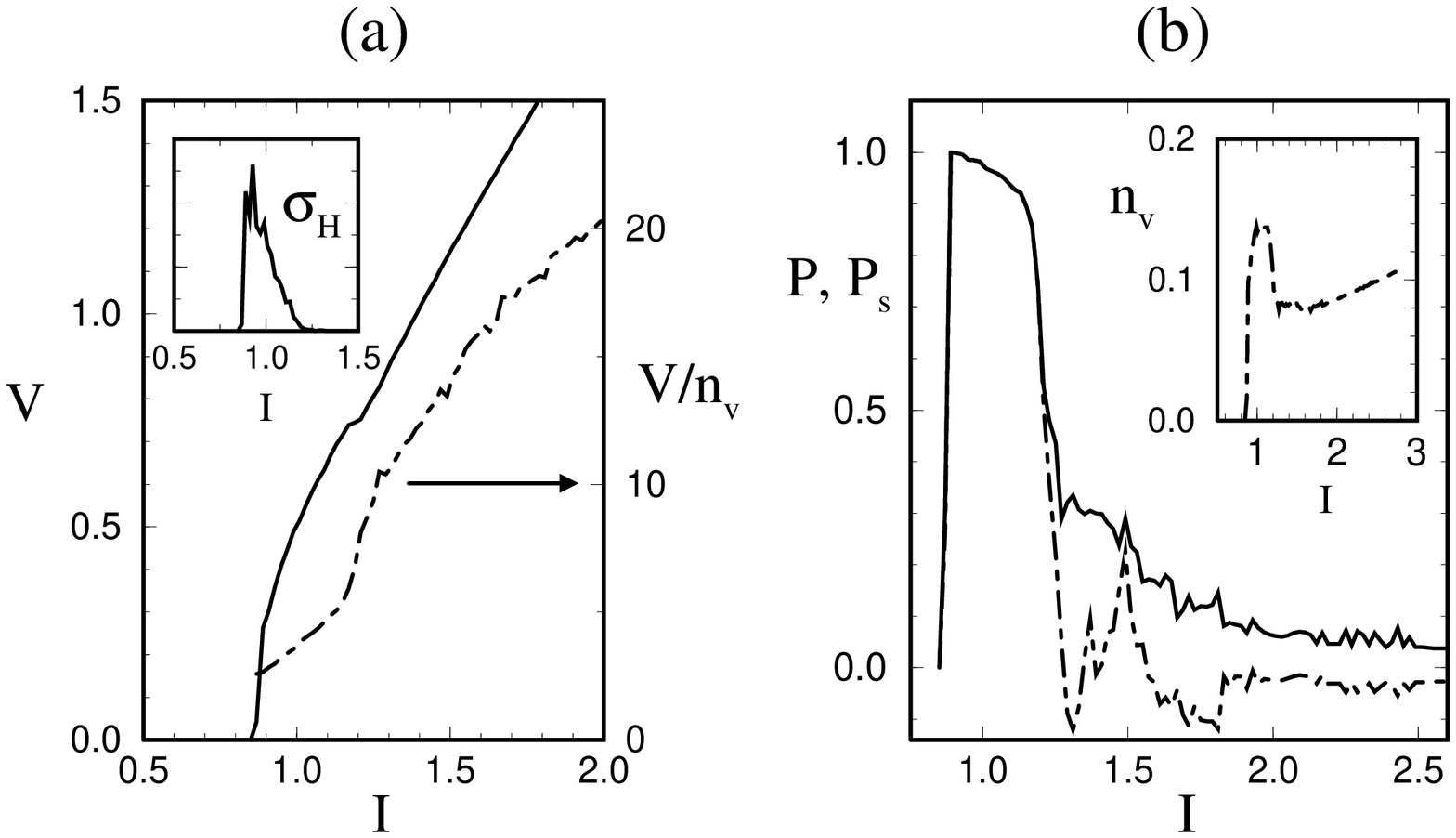}
\end{picture} 
{\small FIG.~6.
Crossover from vortex-dominated to coherent-phase dynamics
as a function of applied current $I$,
probed by  (a) $V$ (full line), $V/n_v$ (dot-dashed) and
the variance $\sigma_H$ of the Hall voltage (inset, in arbitrary units);
(b) $P$ (full line) and $P_s$ (dot-dashed) 
and $n_v$ (inset). All results are obtained in
an upward current sweep of a $20\times20$ array for
$\kappa=0.6$ and $T=0$. For each current value we
used a warm-up time of $1000 t_c$ and an averaging
time of $2000 t_c$.
\label{macros}
}
\end{figure}
\vspace{0.1cm}
\begin{multicols}{2}
\subsection{Crossover from vortex-dominated to coherent-phase dynamics}

In the above we have focused on comparing the LTSEM
images with snapshots of vortex configurations from
numerical simulations. We found
three regions in the $I-V$ characteristics as well as in
the experimental and numerically obtained images.
We related these regions to different types of dynamics. 
The crossover from a vortex-dominated to a
coherent-phase dynamics can be explored
in more detail in the model simulations. 
To this end, we can consider a number 
of quantities that probe the degree of vortex organisation or
the degree of phase coherence.
In particular, we find that the following 
order-parameter-like quantities $P$ and $P_s$ can be used to
distinguish the different current ranges,
\begin{equation}
P\equiv \left\langle \frac{\sum_{Y=1}^{M}
\vert n_c(Y,t)\vert}{N_V(t)}\right\rangle,
\end{equation}
\begin{equation}
P_s\equiv \left\langle \frac{\sum_{Y=1}^{M}
(-1)^Y n_c(Y,t)}{N_V(t)}\right\rangle,
\end{equation}
where
\[
n_c(Y,t)\equiv\sum_{X=1}^{N-1} n({\bf R},t),~
N_V(t)\equiv\sum_{\bf R} \vert n({\bf R},t)\vert.
\]
$X$ and $Y$ denote the $x$ and $y$ component of 
the plaquette coordinate $R$, respectively:
${\bf R}=X{\bf e}_x+Y{\bf e}_y$.
The physical meaning of $P$ and $P_s$ 
can be inferred from the fact that
a vortex that enters at
one side of the array, either leaves the array on the other
side, or is annihilated by an antivortex moving in the opposite
direction. The quantity $n_c(Y,t)$ for row $Y$ 
distinguishes between these two possibilities. In the former
case $n_c$ is nonzero, whereas in the latter case it is, on
average, zero. A value of $P=1$ thus implies that all the vortices
cross the whole array unobstructed, while $P=0$ implies that
the vortices are annihilated in the middle of the array.
The
staggered order parameter $P_s$ measures whether the spatially
resolved vortex dynamics consists of alternating rows of
vortices and antivortices crossing the array. The presence
of domain walls, i.e. two adjacent rows in which vortices
cross the array in the same direction, reduces $P_s$.

On the basis of this interpretation we expect that
Fig. 4(a) (low current region (II)) corresponds to $P\approx 1$,
$P_s\approx 1$; 
and Fig. 4(c) (region (III))
to $P\approx 0$, and $P_s\approx 0$. In Fig.$~6$
we plot $P$, and $P_s$ versus the applied current.
Indeed, directly after depinning, $I>I_c\approx 0.87 N i_c$, 
$P$ and $P_s$ attain
values close to one. As the current increases this value
slowly decreases (low current region (II)).
At $I\approx 1.2 Ni_c$ the value of both order parameters
exhibits a sharp drop
to a much smaller (but {\em nonzero}) value.
In this region the order parameters slowly decrease
to the values near zero in (III). 
Thus, using $P$ and $P_s$ one
can readily establish the type of vortex dynamics
without having to study vortex animations for
each value of the bias current.
In particular, long-time-stable values 
$P\approx 1$, $P_s\approx1$
correspond to a staggered crossing vortex
motion.

We will now correlate the different regions,
mapped out using $P$ and $P_s$,
with the behavior of other quantities. In Fig.
6(a) we show both the voltage and the average
velocity per vortex
$V_v=V/n_v$ (the voltage normalized by the number of vortices)
versus current. 
In the inset
of Fig. $6$(b)
we plot the vortex density.
We observe that
the vortex density displays
a pronounced maximum in  region (II), 
accompanied by a smaller slope of the $V_v(I)$ curve 
as compared to the other regions.
The dynamical properties
of the vortices in this region are thus different
from the other regions.

Vortex jumps across a junction do not only give rise to a contribution
to the longitudinal voltage across the array, but also induce
a fluctuation of the transverse (Hall) voltage around zero.
For the nearly antisymmetric dynamical patterns of vortices in the 
phase-coherence region (III), like the ones shown
in Fig. 4(c), the Hall voltage contributions of the left and right half
of the array tend to cancel by symmetry. In the region with staggered
crossing vortex motion
however, the vortex jumps in the left half of the array do not occur in 
unison with antivortex jumps in the right half, leading to larger
fluctuations in the Hall voltage. Therefore the Hall voltage
 fluctuations may be viewed as a measure of the degree of
 (anti)symmetry in the dynamics. Indeed, as seen in the inset
 of Fig. 6(a),
the magnitude of these fluctuations is reduced dramatically
between the vortex-dominated and the coherent-phase regime.

Figure 6(a) is obtained in an upward current sweep.
We have also performed a downward current sweep from region (III),
starting with a uniform phase configuration. We find that region
(III) is exactly characterized by $P=P_s=0$, whereas in the upward
sweep there are still some asymmetries in the vortex
configuration that yield
small but nonzero values for the order parameters. Entering  the high current end of (II),
$P$ and $P_s$ attain nonzero values. There
is a slight hysteresis in the current value at which $P$ and $P_s$
attain values close to one. 
\begin{figure}
\centerline{
\psfig{figure=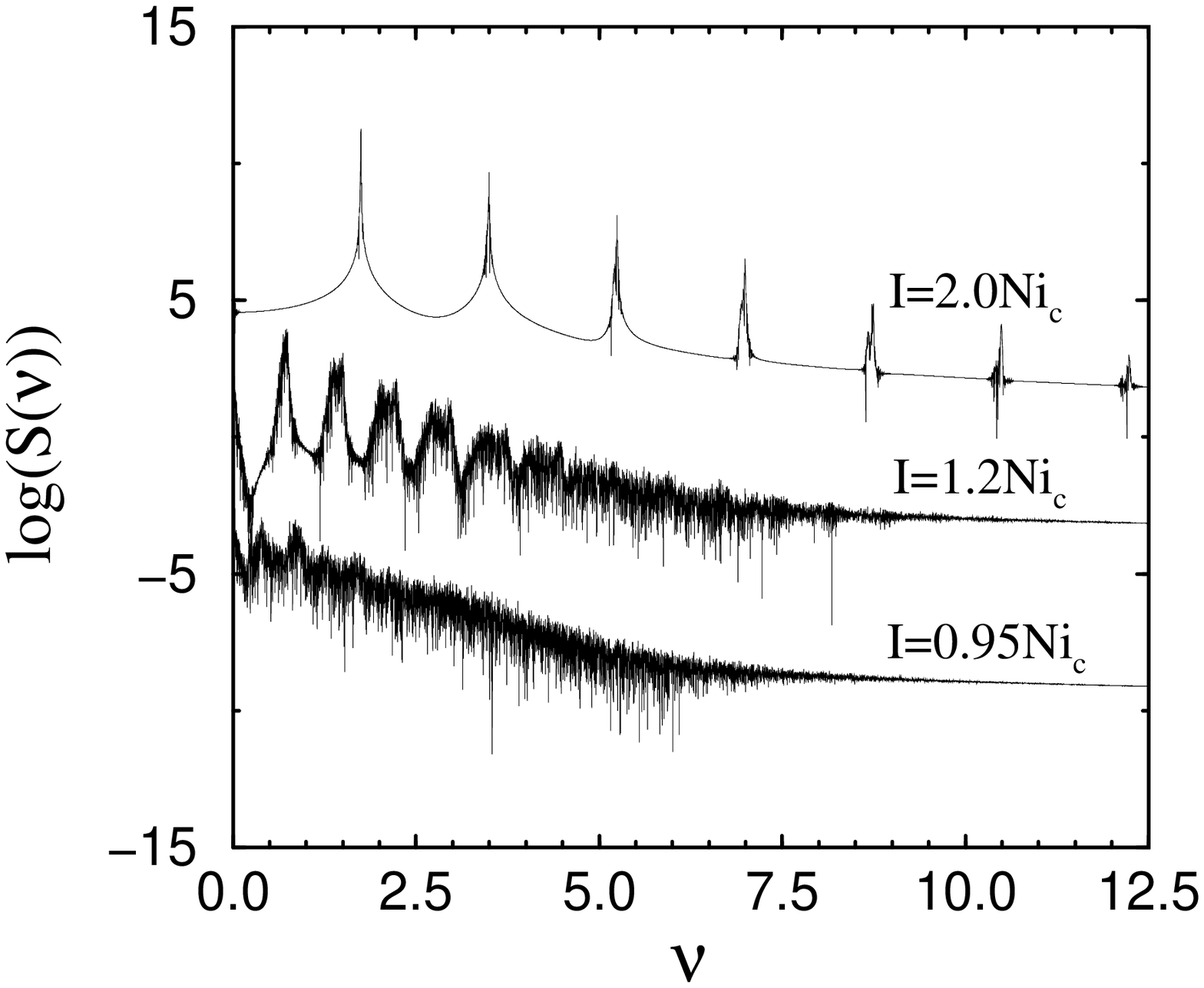,height=7.2cm,width=9.0cm}
}
{\small FIG. 7. 
Calculated power spectrum of the array voltage
for three different bias currents. Subsequent
curves are offset by 5.5 units.
\label{spcomp}
}
\end{figure}
 The power spectrum of the voltage,
 \begin{equation}
 S(\nu)=\left| \int dt~V(t) e^{i2\pi\nu t}\right|^2
 \end{equation}
 is another useful probe for the vortex dynamics. A non-zero voltage can be
 viewed 
 as either created by a vortex jump, or by
 a junction phase slip. When the longitudinal junctions oscillate 
 coherently, $S(\nu)$ consists of sharp peaks  at multiples
 of $\nu=V/(2\pi t_c)$. If the dynamics, however, is dominated
 by incoherent vortex jumps, the peaks in the spectrum are much
 broader. 
 In Fig. 7
 we show $S(\nu)$ for 3 different currents.
 From $i_b= 0.95$, within the ``vortex" region, to $i=2.0$ (region
 (III)) the spectrum changes from a noisy to sharply peaked one.
 These results again illustrate the crossover from a
 vortex-dominated dynamics to
 a coherent-phase dynamics for higher currents.

\section{Conclusions and discussion}

In summary, we have shown that the comparison
of  LTSEM images and results of model calculations
can significantly increase our insight in the dynamics of
Josephson-junction arrays. 
On the one hand, the comparison
of the experimental images with the results of model simulations
corroborates the interpretation of images for 
currents not too far from the array critical current,
and contributes to our understanding of the dynamics underlying
the images outside this current region.
It  shows that the dynamics we found
exists even in the absence
of the disturbance produced by the measuring device.
On the other hand, the agreement with the experimental results
supports the relevance of the model equations employed.
In the context of the samples studied here,
an essential ingredient of these model equations
is the inclusion of (strong)
self-induced magnetic fields.

The successful  comparison between experimental and theoretical results
has enabled us to map out three regions with
different types of dynamics for dc biased arrays with $\lambda_\perp < a$
in zero applied magnetic field.
For bias currents above the array critical current
we found current-induced vortex nucleation at the array edges
parallel to the bias current.
We identified an alternating pattern of crossing vortices  and antivortices
as
the typical vortex dynamics 
existing for bias currents slightly above the array critical
current. We conclude that at least part of the rich structure found in
the experimental $I$-$V$'s is due to the dynamics of 
vortices \cite{BBnote,sbnote}.
For larger currents, the $I-V$ characteristic
becomes linear, and the underlying
dynamics is characterized  by a growing tendency of longitudinal junctions
to oscillate in phase. We have further illustrated  the crossover
from vortex-dominated to coherent-phase like dynamics by numerically 
studying the spectral function, the Hall-voltage fluctuations
and order-parameter-like quantities.

Recently Oppenl\"ander and coworkers \cite{Opp} 
have also calculated I-V characteristics
similar to the experimental ones,
however without taking into account the dynamics 
of the self-induced magnetic fields that are essential
for a successful 
theoretical description of the experimentally observed
microscopic dynamics.
In their work the structure in the differential
resistance is due to the inclusion 
of strong (magnetic) disorder \cite{footnote}. 

In this paper, we have discussed the numerical results
for zero McCumber parameter.
In the experiments, the McCumber parameter was estimated
to be $\beta_c=0.7$.
In our model simulations,
the microscopic dynamics for currents
above but close to the array critical current remains qualitatively
the same for this value of the McCumber parameter.
For higher currents the dynamics is again characterized by a large
degree of spatial coherence.
The region of genuine vortex dynamics shrinks with increasing
$\beta_c$.
For $\beta_c\geq 2.5$ the system enters a row-switched state
immediately above the critical current (for $\lambda_\perp=0.6a$), and  therefore
no vortex-flow regime is found.
For non-zero applied magnetic field,
row-switched states in inductive arrays
were studied in Ref.~\cite{Phi94b}.

For the arrays used in the present studies,
the Kosterlitz-Thouless-Berezinskii phase transition
temperature $T_{KTB}$ is close to the superconducting
transition temperature $T_c$ of the Nb thin films. 
The experiments are performed at temperatures well below
$T_{KTB}$, $\frac{2e k_B T}{\hbar i_c}\sim
1\cdot 10^{-3}$. For these temperatures the effect of thermally
induced vortices on the array dynamics is negligible
\cite{PTthesis}.
We have also performed calculations for $\lambda_\perp\gtrsim a$
(outside the region of the present experiments)
up to $\lambda_\perp=10a$ \cite{Hagfut}.
In this regime we have found
a staggered vortex dynamics similar to the one
observed for $\lambda_\perp\lesssim a$, 
again occuring for currents slightly 
above the array critical current.
This indicates that the staggered vortex dynamics is the generic
dynamics for such currents.

In the LTSEM experiments, the images
corresponding
to the staggered vortex dynamics were found for some  range
of magnetic frustrations around zero. We have simulated 
the dynamics in the lower part of
region (II) (where
the alternating crossing vortex
motion is observed) in a magnetic frustration of
$f=0.01$. For currents slightly above $I_c$, we find that the array evolves
towards a similar state as found for $f=0$,
but now there are on average more vortices with
positive vorticity than  with negative vorticity.\\

\noindent
{\it Acknowledgements.}
We appreciate fabrication of the samples by P. A. A. Booi and
S. P. Benz, National Institute of Standards and Technology,
Boulder (CO).
We thank G. Filatrella, C. Giovannella, W. G\"uttinger,
F. Hilbert, D. Hoffmann,
M. Keck, A. Laub, J. Oppenl\"ander, T. Tr\"auble,
and A. V. Ustinov
for valuable discussions.
This work was supported by the EU Human Capital and Mobility
program (contract No. CHRX-CT92-0068) and the Bundesminister
f\"ur Bildung, Wissenschaft, Forschung, und Technologie
under grant No.~13N6436.
This work has
been partially supported by NSF grant DMR-9521845 (JVJ),
by the Dutch
organisation for fundamental research (FOM) (PT,JEH), and by
the Bayerischer Forschungsverbund 
Hochtemperatur-Supraleiter (FORSUPRA) (TJH).

\end{multicols}

\begin{references}
\bibitem{sbnote} 
In the context of superconducting bridges,
several authors (see, for example, Refs.~\cite{Hue74,Ped})
have reported structures in the I-V characteristic 
and attributed those to viscous vortex motion.
\bibitem{footnote}  
In Ref. \cite{Opp} it is assumed that there are no transverse voltage
fluctuations, i.e. the supercurrents in the transverse
direction
are held constant.
As a result, the array dynamics decouples into $N$ independent 
one-dimensional, non-uniform and nonlinear chain equations 
for each column. The nontrivial nature of the 
couplings of the two-dimensional system is retained in the
randomly chosen, time-independent phase-differences 
in the transverse direction,
that  one may interpret
in terms of the presence of a
(strong) static random magnetic field.
\end{references}
\end{document}